\documentclass[aps,prl,twocolumn,amsmath,showpacs,letterpaper,floatfix]{revtex4-1}

\newcommand{\bra}[1]{\left\langle #1\right|}
\newcommand{\ket}[1]{\left| #1\right\rangle}

\usepackage{amssymb,amsmath,amsthm,amsfonts,bm}
\usepackage{graphicx}
\usepackage{epsfig}
\usepackage{subfigure}
\usepackage{amsmath}			
\usepackage{grffile}
\usepackage{color}
\newtheorem*{definition}{Definition}

\begin{document}

\title{Quantum Process Nonclassicality}
\author{Saleh Rahimi-Keshari}
\email{s.rahimik@gmail.com}
\affiliation{Centre for Quantum Computation and Communication Technology, School of Mathematics and Physics, University of Queensland, Brisbane,   QLD 4072, Australia}

\author{Thomas Kiesel and Werner Vogel}
\affiliation{Arbeitsgruppe Quantenoptik, Institut f\"ur Physik, Universit\"at Rostock, D-18051 Rostock, Germany}

\author{Samuele Grandi}
\affiliation{Dipartimento di Fisica dell'Universit\`{a} degli Studi di Milano, I-20133 Milano, Italy}

\author{Alessandro Zavatta and Marco Bellini}
\affiliation{Istituto Nazionale di Ottica, INO-CNR, Largo Enrico Fermi, 6, I-50125 Firenze, Italy}
\affiliation{Department of Physics and LENS, University of Firenze, 50019 Sesto Fiorentino, Firenze, Italy}

\begin{abstract}
We propose a definition of nonclassicality for a single-mode quantum-optical process based on its action on coherent states. 
If a quantum process transforms a coherent state to a nonclassical state, it is verified to be nonclassical. To identify nonclassical processes, we introduce a representation for quantum processes, called the process-nonclassicality quasiprobability distribution, whose negativities indicate nonclassicality of the process. Using this distribution, we derive a relation for predicting nonclassicality of the output states for a given input state.
We experimentally demonstrate our method by considering  the single-photon addition as a nonclassical process, and predicting nonclassicality of the output state for an input thermal state.
\end{abstract}

\maketitle
\paragraph{Introduction.}
The ability of detecting any nonclassicality generated by any quantum device enables us to manipulate and control the evolution of a quantum state. In particular, this plays a central role in implementation of quantum information processing and communication~\cite{Nielsen-Chuang}.

In general, the problem of the characterization of an unknown quantum device is addressed by means of quantum process tomography \cite{PCZ97,DL01, MRL08}. A general method for quantum process tomography was recently proposed that is based on probing a quantum process (described by a completely positive and linear map $\mathcal{E}$) using coherent states to characterize the process tensor in the Fock basis, with a fixed maximum number of photons~\cite{lobino2008,rahimi-keshari2011,LobinoQPT2,lvovsky2012}.
However, any photon-number cutoff will transform a classical state into a nonclassical one, as a finite sum of nonclassical states is always nonclassical. 
Therefore, the previously known methods are not able to distinguish quantum processes whose outputs are classical for any classical input state from those that may convert classical input states into nonclassical output states.

For this purpose a universal nonclassicality test of the output states is indispensable. Nonclassicality of quantum states is characterized by the Glauber-Sudarshan representation \cite{Glauber,Sudarshan} of the density operator $\hat \rho$,
\begin{equation}
\hat \rho=\int \text{d}^2 \alpha P(\alpha) \ket{\alpha}\bra{\alpha}.
\label{Glauber-Sudarshan}
\end{equation} 
If the $P$ function has the form of a classical probability density, the corresponding quantum state is said to have classical analogue \cite{Titulaer}, otherwise the state is referred to as nonclassical~\cite{Mandel}. 
However, in practice the $P$~function is highly singular for many quantum states,
so that it cannot be used to experimentally check the nonclassicality in general.

A recently proposed method for verifying nonclassicality of quantum states is to use a regularized version of the $P$ function, referred to as the nonclassicality quasiprobability distribution (NQD), $P_\Omega(\beta)$. Its negativities indicate the nonclassicality of any quantum state~\cite{Kiesel}, and it can be directly sampled by homodyne detection \cite{Kiesel-POmega-Spats,Kiesel-POmega-Squeeze}. The relation between the NQD and the $P$~function is easily formulated by their Fourier transforms, i.e., the characteristic functions $\Phi_\Omega(\xi)$ and $\Phi(\xi)$, respectively. The function $\Phi_\Omega$ is obtained by multiplying $\Phi$ with a proper filter function, for a detailed discussion of the requirements we refer to~\cite{Kiesel}.
An example of such a filter function is
\begin{equation}
\Omega_w(\xi)=\frac{1}{\mathcal N} \int \text{d}^2\eta e^{-|\eta|^4} e^{-|\frac{\xi}{w} + \eta|^4} \ ,
\label{nonclas-filter}
\end{equation}
where $\mathcal N$ ensures the normalization $\Omega_w(0){=}1$.

In this Letter, we propose a definition of nonclassicality for single-mode quantum-optical processes. We introduce a method for detecting nonclassical processes by testing the nonclassicality of the output states for coherent states at the input.
If there exists an input coherent state leading to a nonclassical output state, the quantum process is nonclassical.
This method enables us to identify nonclassical quantum processes that may transfer classical input states into nonclassical output states.
We derive a relation for predicting the NQD of the output state for given input states. Moreover,
we experimentally demonstrate our method by verifying  
the single-photon addition to be a nonclassical process and predicting nonclassicality of the output state for an input thermal state.

\paragraph{Nonclassical processes.}

For a general input quantum state $\hat \rho$, the output of a quantum process described by the map $\mathcal{E}(\hat \rho)$ is obtained 
by using Eq.~(\ref{Glauber-Sudarshan}) and the linearity of the map, 
\begin{equation}
\mathcal{E}(\hat \rho)=\int \text{d}^2 \alpha P_{\text{in}}(\alpha) %
\mathcal{E}(\ket{\alpha}\bra{\alpha})
\label{proc-out}
\end{equation}  
where $P_{\text{in}}(\alpha)$ is the $P$ function of the input state. As the map may not be trace-preserving in general, the output state, $\hat{\rho}_{\mbox{\tiny $\mathcal{E}$}} \propto \mathcal{E}(\hat{\rho}) $, is obtained from this expression simply by normalization. 
The $P$ function of the output state is given by
\begin{equation}
P_{\text{out}}(\beta) = \int \text{d}^2 \alpha P_{\text{in}}(\alpha) P_{\mathcal{E}}(\beta |\alpha),
\label{P-in-out}
\end{equation}
where $P_{\mathcal{E}}(\beta | \alpha)$ is the $P$ function of the output state of the process conditioned on the input state being the coherent state $|\alpha\rangle$.

From \eqref{P-in-out} it follows that if the output states of a quantum process for input coherent states are classical states, i.e., having $P$~functions that are positive semidefinite, then the output of the process for any classical input quantum states
will be always a classical state. This motivates us to define nonclassicality of quantum processes as follows.
\begin{definition}
A quantum process is nonclassical if it transforms an input coherent state to a nonclassical state.
\end{definition}

Therefore, based on this definition, a {\em classical process} transforms all coherent states to classical states, and the output state is classical for any classical input state. Also, nonclassicality of the output state for only one coherent state is sufficient evidence that the process is nonclassical.

As the regularized version of the $P$ function, $P_\Omega(\alpha)$, is an appropriate representation of quantum states for experimentally verifying nonclassicality, 
we introduce a related characterization of quantum processes.
The regularized version of $P_{\mathcal E}(\beta | \alpha)$, denoted as $P_{\Omega,\mathcal{E}}(\beta | \alpha)$, a regular function of two complex variables,
is a representation of the process that we use to verify its nonclassicality. This conditioned quasiprobability distribution is denoted as the process-nonclassicality quasiprobability distribution (PNQD), which unambiguously identifies the nonclassicality of a given quantum process. For a nonclassical process there exists an input coherent state $\ket{\alpha_0}$ and $\beta_0$ such that $P_{\Omega,\mathcal{E}}(\beta_0 |\alpha_0){<}0$.

We note that the nonclassicality of a quantum process does not imply that the output state is nonclassical for any classical input states, as we shall see in the following example of the cat-generation process. Having knowledge of the PNQD, $P_{\Omega,\mathcal{E}}(\beta | \alpha)$, and using \eqref{P-in-out}, one can find the NQD of the output state for an input state described by $P_{\text{in}}(\alpha)$ via
\begin{equation}
P_{\Omega}(\beta)= \int \text{d}^2\alpha P_{\text{in}}(\alpha) P_{\Omega,\mathcal{E}}(\beta | \alpha) \ .
\label{nonclas-quas-in-out}
\end{equation}
In case the $P$ function of the input state is highly singular, by using the Parseval identity the NQD of the output state can be obtained as
\begin{equation}
P_{\Omega}(\beta)= \int \text{d}^2\nu \Phi_{\text{in}}(\nu) \tilde{P}_{\Omega,\mathcal{E}}(\beta | \nu) \ ,
\end{equation}
where $\Phi_{\text{in}}(\nu)$ is the characteristic function of the $P$ function and $\tilde{P}_{\Omega,\mathcal{E}}(\beta | \nu)$ is the Fourier transform of $P_{\Omega,\mathcal{E}}(\beta |\alpha)$.

For unknown quantum processes the PNQD can be estimated by sampling the NQD of the output states~\cite{Kiesel-POmega-Spats,Kiesel-POmega-Squeeze} for a sufficiently large number of input coherent states.  In principle, for any unknown quantum process, $P_{\Omega,\mathcal{E}}(\beta|\alpha)$ can even be uniquely determined by knowing the action of the process on only an arbitrary small compact set of input coherent states $\ket\alpha$~\cite{Supplement}.

\begin{figure}[h]
\includegraphics[width=0.8\columnwidth]{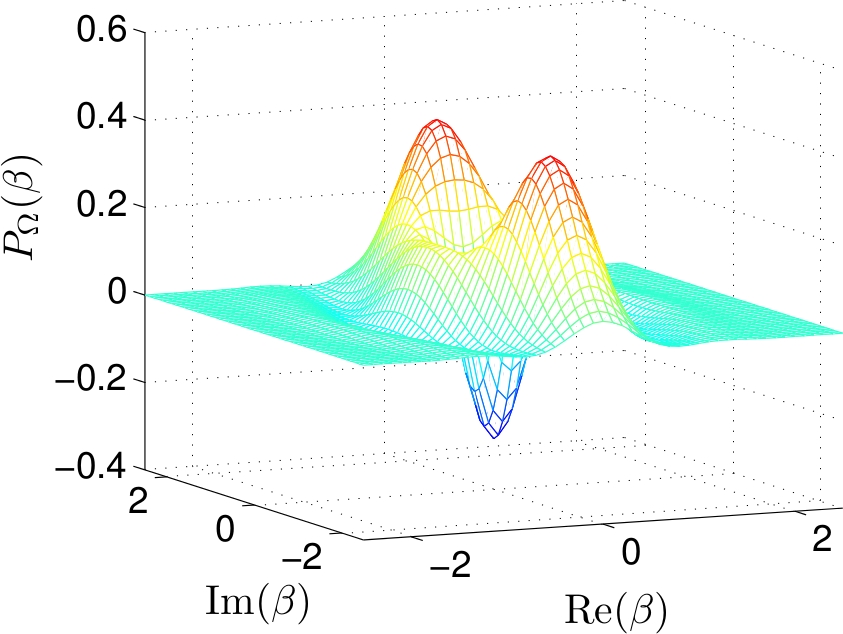}
\caption{(Color online) The NQD  for the output of the single-photon subtraction process, acting on a squeezed vacuum state with variances $V_x {=} 0.5$ and $V_p {=} 3.0$. We used the filter function~\eqref{nonclas-filter} with the filter width $w {=} 1.5$. The negativities indicate the nonclassicality of the state.}
\label{fig:POmega1}
\end{figure}

\paragraph{Examples of classical and nonclassical processes.}
Examples of classical processes include the photon subtraction and the interaction of a state with a thermal bath; for further details on classical maps, cf. Ref.~\cite{Gehrke}. For the photon-subtraction process, the map
\begin{equation}
\mathcal{E}(\ket{\alpha}\bra{\alpha})
=\hat{a} \ket{\alpha}\bra{\alpha} \hat{a}^{\dagger} = |\alpha |^2 \ket{\alpha}\bra{\alpha} 
\end{equation}
yields a classical state. Consequently, the output will be classical for any classical input state~\cite{zavatta2008}. However, a nonclassical input state may be transformed to an output state with modified nonclassical properties~\cite{grangier2006}, see Fig.~\ref{fig:POmega1}. Hence the output state of a classical process for a nonclassical input state may not be classical.

Let us consider a model for the process of decoherence caused by a thermal bath of mean occupation number $\bar{n}$ \cite{Richter-Vogel},  
the characteristic function of the $P$ function of the state at time $t$ is given by \cite{Marian}
\begin{equation}
\Phi(\xi,t)= \exp\left[-|\xi|^2 (\bar{n}-(\bar{n}+1)e^{-2\gamma t})\right] \Phi_Q(\xi e^{-\gamma t},0)  \ ,
\end{equation}
where $\Phi_Q(\xi,0)$ ($\Phi_Q(\xi,0){=}\exp[-|\xi|^2]\Phi(\xi,0)$) is the characteristic function of the $Q$~function of the state at time $t{=}0$, $t$ is the interaction time, and $\gamma$ is the damping rate. The $Q$~function of any quantum state is a positive semidefinite function~\cite{perelomov}. Hence, for $\frac{\bar{n}}{\bar{n}+1}{>}e^{-2\gamma t}$ the $P$~function is positive semidefinite, as it is given by the convolution of two positive semidefinite functions, and the output state for any initial nonclassical state is always classical. In this case, this process is classical.

\begin{figure}[h]
\includegraphics[width=0.8\columnwidth]{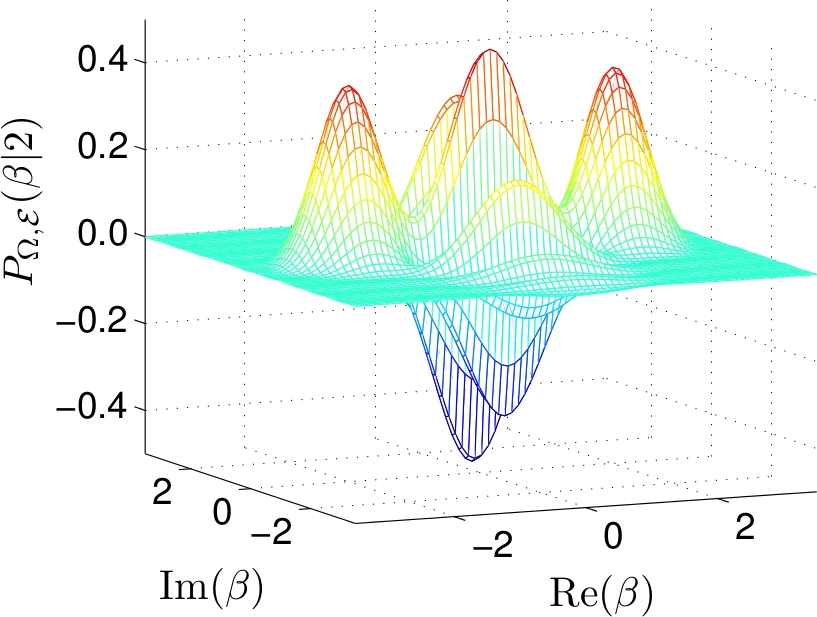}
\caption{(Color online) The PNQD of the cat-generation process, for an input coherent state with $\alpha{=}2.0$. The filter function is the same as in Fig.~\ref{fig:POmega1}.}
\label{fig:POmega1b}
\end{figure}

An example of a nonclassical process is the cat-generation process. The unitary evolution associated with the Hamiltonian $\hat{H}_{\text{Kerr}}{=}\chi(\hat{a}^{\dagger}\hat{a})^2$, generates the Schr\"odinger cat state at time $t{=}\frac{\pi}{2\chi}$ ($\hbar{=}1$) \cite{Milburn,YS86}
\begin{align}
\mathcal{E}(\ket{\alpha}\bra{\alpha})&=e^{-i\frac{\pi}{2}(\hat{a}^{\dagger}\hat{a})^2}\ket{\alpha}\bra{\alpha}e^{i\frac{\pi}{2}(\hat{a}^{\dagger}\hat{a})^2}\nonumber \\
&=\frac{1}{2}(\ket{\alpha}+i\ket{-\alpha})(\bra{\alpha}-i\bra{-\alpha}) \ .
\end{align}
This is a nonclassical process, as the PNQD takes on negative values; see Fig.~\ref{fig:POmega1b}. As would be expected, this nonclassical process converts a classical state to a nonclassical one.
However, for certain classical input states the output state can be classical. As the corresponding unitary operator is a function of the photon number operator $\hat{n}{=}\hat{a}^\dagger\hat {a}$, it leaves the photon number states $|n\rangle \langle n|$ unchanged. As a consequence, any statistical mixture of photon number states remains unchanged.  Therefore, the output state of this process for an input thermal state is the same thermal state, which yields a classical output from a nonclassical process with classical input.  

Last but not least, even a nonclassical process with nonclassical input can have a classical output state. A simple example is the application of the squeezing operation on a squeezed input state. For example, when a squeezed vacuum input state is squeezed again with the same amount of squeezing but 
in the quadrature orthogonal to the original squeezing, the squeezed state transforms into the vacuum state.

\paragraph{Experimental demonstration of a nonclassical process.}
Based on our definition, the single-photon addition process is a nonclassical process, as it transforms the vacuum state to the single-photon state, i.e., a coherent state with zero amplitude to a nonclassical state. In the following, we experimentally demonstrate our method by applying it to the single-photon addition process.

Stimulated parametric down-conversion is used to generate the single-photon-added coherent states~\cite{zavatta04:science,zavatta05}. The core of the experimental setup is a $\chi^{(2)}$-nonlinear crystal [$\beta$-barium borate (BBO), type I] pumped by a 90-mW UV beam obtained by frequency doubling 1.5-ps pulses at 785 nm from a mode-locked Ti:sapphire laser. The spontaneous parametric down-conversion from the crystal consists in pairs of entangled photons emitted in two well-defined directions called signal and idler. When a seed coherent state is injected in the crystal along the signal direction, stimulated emission also takes place.
A single-photon avalanche silicon detector is placed along the idler beam after spatial and spectral filtering. A click from this detector heralds the generation of a single-photon-added
coherent state in a well-defined spatiotemporal signal mode, which is then characterized by time-domain homodyne detection~\cite{zavatta02:josab}.
In the experiment, we analyzed the photon-added states with 13 different input coherent-state amplitudes. For each acquisition the homodyne phase was varied between zero and $\pi$ and actively locked to 10 different values by monitoring the DC level from the homodyne receiver.

\begin{figure*}[t]
	\includegraphics[width=0.32\textwidth]{{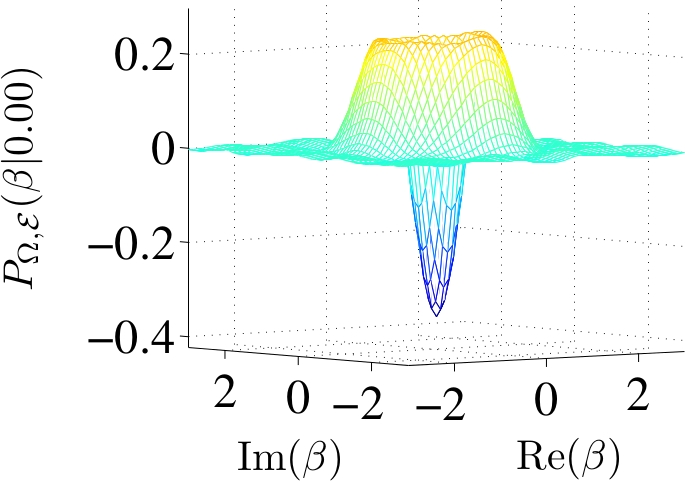}}
	\includegraphics[width=0.32\textwidth]{{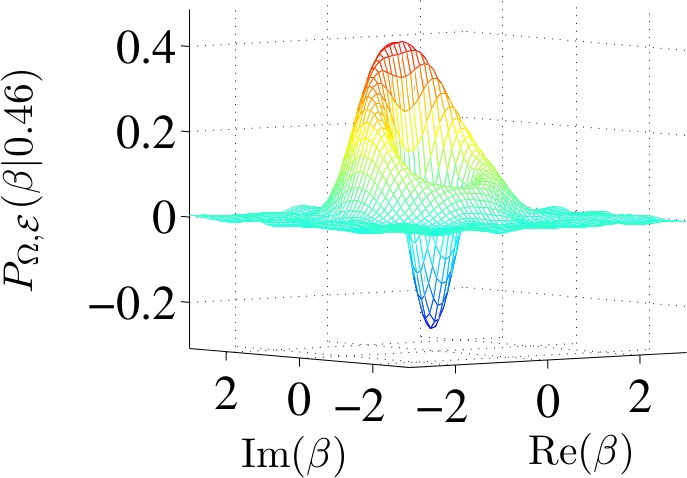}}
	\includegraphics[width=0.32\textwidth]{{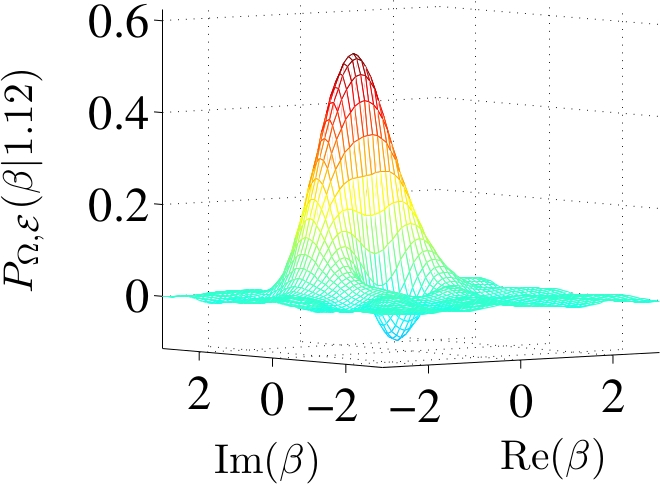}}
	\caption{(Color online) PNQD of the single-photon-addition process with $w{=}1.2$ at different amplitudes $\alpha{=}0.00$, $\alpha{=}0.46$ and $\alpha{=}1.12$ of input coherent state (from left to right).}
	\label{fig:nqp:spacs}
\end{figure*}
\begin{figure}[h]
	\includegraphics[width=\columnwidth]{{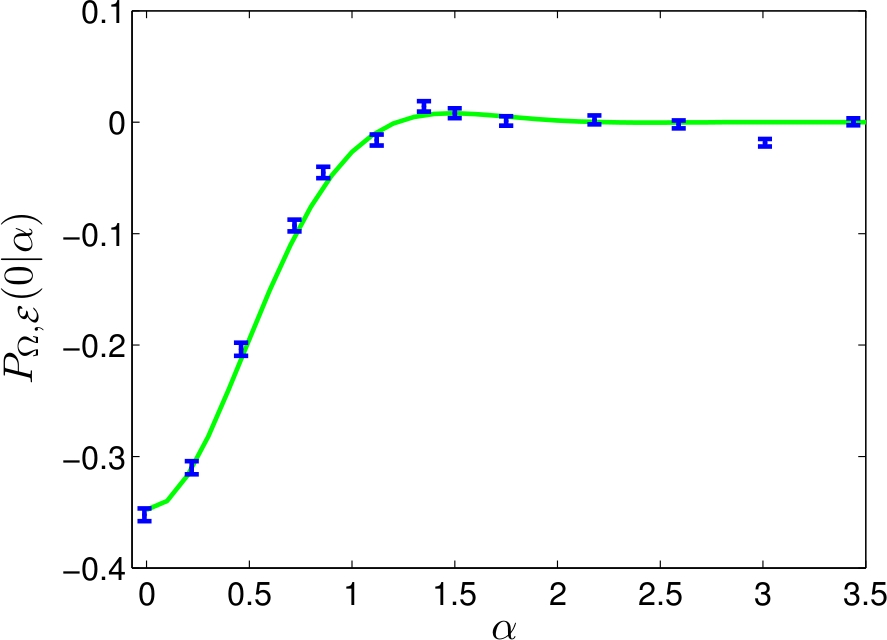}}
	\caption{(Color online) PNQD with $w{=}1.2$ for different input amplitudes $\alpha$ for the single-photon-addition process, evaluated at $\beta {=} 0$. The error bars correspond to one standard deviation. The green solid line represents the theoretical expectation.}
	\label{fig:nqp:spacs:cross}
\end{figure}

 
Now we prove experimentally that this process is a nonclassical one.
To estimate the PNQD of this process from experimentally recorded quadrature distributions, we use the sampling approach which has already been applied to determine the NQD in \cite{Kiesel-POmega-Squeeze}. 
The PNQD is reconstructed by using the filter function~\eqref{nonclas-filter} with the filter width $w{=}1.2$. The effect of the quantum efficiency is removed as described in~\cite{Supplement}. The obtained results are shown for three different input coherent states in Fig.~\ref{fig:nqp:spacs}. We observe negativities for different amplitudes of the coherent input state $\ket{\alpha}$, with decreasing negativity for increasing $\alpha$. Obviously, the negativity appears close to the origin of phase space, i.e.,~at $\beta{=}0$. Therefore, we examine the dependence of the PNQD on the input amplitude $\alpha$ at $\beta{=}0$ more closely; see Fig.~\ref{fig:nqp:spacs:cross}. It is clearly seen that the negativity is statistically significant for low input amplitudes $\alpha$, which eventually yields the sought experimental proof of the nonclassicality of the process. 
For larger amplitudes, the negativity vanishes at $\beta{=}0$. However, this does not mean that the output state for an input coherent state with large amplitude is classical. As the single-photon-added coherent states are nonclassical for any input amplitude \cite{zavatta05}, one will find negativities of the PNQD at values of $\beta$ different from zero.

By using the experimentally estimated PNQD for the single-photon-addition process, we are able to estimate the NQD of the output state  via Eq. (5) for a thermal input state with low mean photon number.  The fact that photon addition is probabilistic is properly taken into account; for details see \cite{Supplement}. In Fig.~\ref{fig:POmega:SPATS} we show the predicted NQD of the output state for a thermal input state, displaying strongly significant negativities, which prove nonclassicality of the output. This estimate of the  NQD of the output state is in good agreement with the directly measured NQD of single-photon-added thermal states~\cite{Kiesel-POmega-Spats}.

\begin{figure}[h]
	\includegraphics[width=\columnwidth]{{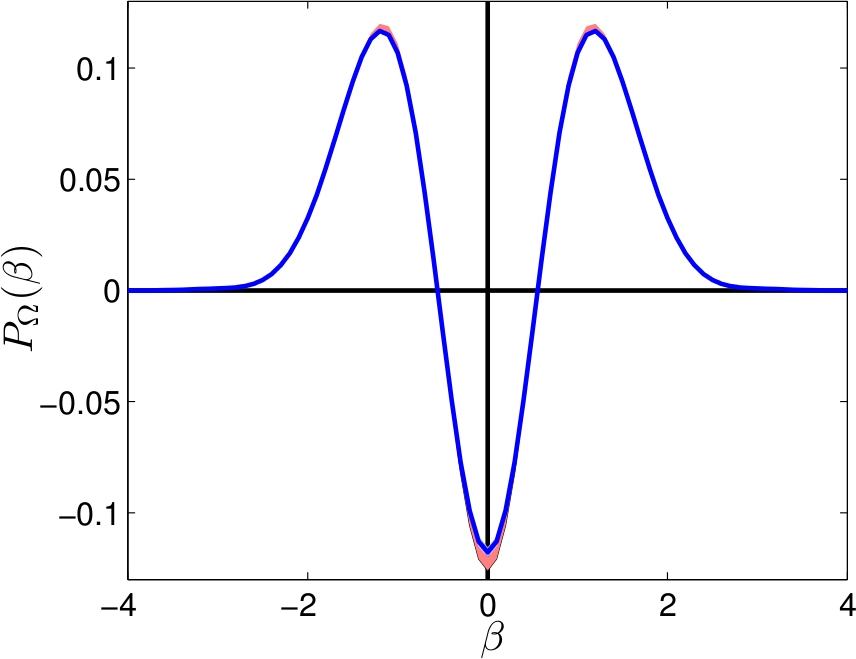}}
	\caption{(Color online) Estimated NQD of the output state of the single-photon addition process	for a thermal input state with mean thermal photon number $\bar n {=} 0.5$. The standard deviation, shown by the blue shaded area, is mostly hidden by the linewidth, the systematic error is displayed by the red shaded area, see \cite{Supplement}.}
	\label{fig:POmega:SPATS}
\end{figure}

\paragraph{Conclusions.}
We have proposed a definition of nonclassicality of a quantum process through its action on coherent states. Based on this definition, any quantum process that transforms a coherent state to a nonclassical one is identified as a nonclassical process, which may transfer a classical state to a nonclassical one. For classical processes the output state is guaranteed to be classical for any input classical states. 
A classical process can also be useful to transform the nonclassical properties of the input state into another form, which is desired for some applications.

Nonclassical processes are necessary for the generation of nonclassical states, and subsequently they can be used to create entanglement by overlapping them on a beam splitter~\cite{kim2002, wang2002}. Conversely, interference of classical states will not generate entanglement. The presented method enables us to check whether an unknown quantum device can generate nonclassical states and to predict nonclassicality of the output state. One useful application for the method is to verify nonclassicality (classicality) in multimode quantum devices and channels, by considering each input to output connection as a single-mode process, when output states are required to be entangled (unentangled) states.

\paragraph{Acknowledgments.}
We thank Jan Sperling and Tim Ralph for helpful discussions. This work was supported by the Deutsche Forschungsgemeinschaft through SFB 652.
S.R. also acknowledges support from the Australian Research Council Centre of Excellence for Quantum Computation and Communication Technology
Project No. CE110001027).
A.Z. and M.B. acknowledge support of Ente Cassa di Risparmio di Firenze, and the EU under ERA-NET CHIST-ERA project QSCALE.


\begin{thebibliography}{}

\bibitem{Nielsen-Chuang} M. A. Nielsen and I. L. Chuang, \textit{Quantum Computation and Quantum Information} (Cambridge University Press, Cambridge, England, 2000).
\bibitem{PCZ97} J. F. Poyatos, J. I. Cirac, and P. Zoller, Phys. Rev. Lett. \textbf{78}, 390 (1997).
\bibitem{DL01} G. M. D'Ariano and P. Lo Presti, Phys. Rev. Lett. \textbf{86}, 4195 (2001).
\bibitem{MRL08} M. Mohseni, A. T. Rezakhani, and D. A. Lidar, Phys. Rev. A \textbf{77}, 032322 (2008).
\bibitem{lobino2008} M. Lobino, D. Korystov, C. Kupchak, E. Figueroa, B. C. Sanders, and A. I. Lvovsky, Science {\bf 322}, 563 (2008).
\bibitem{rahimi-keshari2011} S. Rahimi-Keshari, A. Scherer, A. Mann, A. T. Rezakhani, A. I. Lvovsky, and B. C. Sanders, New J. Phys. {\bf 13}, 013006 (2011).
\bibitem{LobinoQPT2} M. Lobino, C. Kupchak, E. Figueroa, and A. I. Lvovsky,
Phys. Rev. Lett. \textbf{102}, 203601 (2009).
\bibitem{lvovsky2012}R. Kumar, E. Barrios, C. Kupchak, and A. I. Lvovsky, Phys. Rev. Lett. {\bf 110}, 130403 (2013).
\bibitem{Glauber} R. J. Glauber, Phys. Rev. \textbf{131}, 2766 (1963).
\bibitem{Sudarshan} E. C. G. Sudarshan, Phys. Rev. Lett. \textbf{10}, 277 (1963).
\bibitem{Titulaer} U. M. Titulaer and R. J. Glauber, Phys. Rev. {\bf 140}, B676 (1965).
\bibitem{Mandel} L. Mandel, Phys. Scr. {\bf T12}, 34 (1986).
\bibitem{Kiesel} T. Kiesel and W. Vogel, Phys. Rev. A \textbf{82}, 032107 (2010).
\bibitem{Kiesel-POmega-Spats} T. Kiesel, W. Vogel, M. Bellini, and A. Zavatta, Phys. Rev. A \textbf{83}, 032116 (2011).
\bibitem{Kiesel-POmega-Squeeze} T. Kiesel, W. Vogel, B. Hage, and R. Schnabel, Phys. Rev. Lett. \textbf{107}, 113604 (2011). 
\bibitem{Supplement} See supplementary material.
\bibitem{Gehrke} C. Gehrke, J. Sperling, and W.Vogel,
      Phys. Rev. A {\bf 86}, 052118 (2012).
\bibitem{zavatta2008} A.~Zavatta, V.~Parigi, M.~S.~Kim, and M.~Bellini, New J. Phys. \textbf{10}, 123006 (2008).
\bibitem{grangier2006} A.~Ourjoumtsev, R.~Tualle-Brouri, J.~Laurat, and P.~Grangier, Science \textbf{312}, 83 (2006).
\bibitem{Richter-Vogel} T. Richter and W. Vogel, Phys. Rev. A \textbf{76}, 053835 (2007).
\bibitem{Marian} P. Marian and T. A. Marian, J. Phys. A \textbf{33}, 3595 (2000).
\bibitem{perelomov} A. M. Perelomov, \textit{Generalized Coherent States and Their Applications} (Springer-Verlag, Berlin, 1986).
\bibitem{Milburn} G. J. Milburn, Phys. Rev. A \textbf{33}, 674 (1986).
\bibitem{YS86} B. Yurke and D. Stoler, Phys. Rev. Lett. \textbf{57}, 13 (1986).

\expandafter\ifx\csname natexlab\endcsname\relax\def\natexlab#1{#1}\fi
\expandafter\ifx\csname bibnamefont\endcsname\relax
  \def\bibnamefont#1{#1}\fi
\expandafter\ifx\csname bibfnamefont\endcsname\relax
  \def\bibfnamefont#1{#1}\fi
\expandafter\ifx\csname citenamefont\endcsname\relax
  \def\citenamefont#1{#1}\fi
\expandafter\ifx\csname url\endcsname\relax
  \def\url#1{\texttt{#1}}\fi
\expandafter\ifx\csname urlprefix\endcsname\relax\def\urlprefix{URL }\fi
\providecommand{\bibinfo}[2]{#2}
\providecommand{\eprint}[2][]{\url{#2}}

\bibitem[{\citenamefont{Zavatta et~al.}(2004)\citenamefont{Zavatta, Viciani,
  and Bellini}}]{zavatta04:science}
\bibinfo{author}{\bibfnamefont{A.}~\bibnamefont{Zavatta}},
  \bibinfo{author}{\bibfnamefont{S.}~\bibnamefont{Viciani}}, \bibnamefont{and}
  \bibinfo{author}{\bibfnamefont{M.}~\bibnamefont{Bellini}},
  \bibinfo{journal}{Science} \textbf{\bibinfo{volume}{306}},
  \bibinfo{pages}{660} (\bibinfo{year}{2004}).


\bibitem[{\citenamefont{Zavatta et~al.}(2005)\citenamefont{Zavatta, Viciani,
  and Bellini}}]{zavatta05}
\bibinfo{author}{\bibfnamefont{A.}~\bibnamefont{Zavatta}},
  \bibinfo{author}{\bibfnamefont{S.}~\bibnamefont{Viciani}}, \bibnamefont{and}
  \bibinfo{author}{\bibfnamefont{M.}~\bibnamefont{Bellini}},
  \bibinfo{journal}{Phys.\ Rev.\ A} \textbf{\bibinfo{volume}{72}},
  \bibinfo{pages}{023820} (\bibinfo{year}{2005}).

\bibitem[{\citenamefont{Zavatta et~al.}(2002)\citenamefont{Zavatta, Bellini,
  Ramazza, Marin, and Arecchi}}]{zavatta02:josab}
\bibinfo{author}{\bibfnamefont{A.}~\bibnamefont{Zavatta}},
  \bibinfo{author}{\bibfnamefont{M.}~\bibnamefont{Bellini}},
  \bibinfo{author}{\bibfnamefont{P.~L.} \bibnamefont{Ramazza}},
  \bibinfo{author}{\bibfnamefont{F.}~\bibnamefont{Marin}}, \bibnamefont{and}
  \bibinfo{author}{\bibfnamefont{F.~T.} \bibnamefont{Arecchi}},
  \bibinfo{journal}{J.\ Opt.\ Soc.\ Am.\ B} \textbf{\bibinfo{volume}{19}},
  \bibinfo{pages}{1189} (\bibinfo{year}{2002}).

\bibitem{kim2002} M.~S.~Kim, W.~Son, V.~Bu{\v z}ek, and P.~L.~Knight, Phys. Rev. A \textbf{65}, 032323 (2002).
\bibitem{wang2002} Wang Xiang-bin, Phys. Rev. A \textbf{66}, 024303 (2002). 

\end{thebibliography}

\begin{thebibliography}{99}

\bibitem{UniNonclWitness} T. Kiesel and W. Vogel, Phys. Rev. A. {\bf 85}, 062106 (2012).
\bibitem{Kraus} K. Kraus, Ann. Phys. \textbf{64}, 311 (1971).

\bibitem{MW} L. Mandel and E. Wolf, \textit{Optical Coherence and Quantum Optics} (New York: Cambridge University Press).
\bibitem{Cahill} K. E. Cahill, Phys. Rev. \textbf{138}, B1566 (1965).

\bibitem{NonclQuasiProbSqueeze} T. Kiesel, W. Vogel, B. Hage, and R. Schnabel,
		Phys. Rev. Lett. {\bf 107}, 113604 (2011).
\bibitem {NonclQuasiProbSPATS} T. Kiesel, W. Vogel, M. Bellini, and A. Zavatta,
		Phys. Rev. A {\bf 83}, 032116 (2011).

\bibitem{NonclWitnessSimulation} T. Kiesel and W. Vogel,
		\pra {\bf 86}, 032119 (2012).


\bibitem{lvovsky2012}R. Kumar, E. Barrios, C. Kupchak, and A. I. Lvovsky, Phys. Rev. Lett. {\bf 110}, 130403 (2013).



\end{thebibliography}

\clearpage
\begin{widetext}
\begin{center}

\medskip

{\large{\bf Supplementary Material: Quantum Process Nonclassicality}}

\medskip

Saleh Rahimi-Keshari$^*$, Thomas Kiesel, Werner Vogel, Samuele Grandi, Alessandro Zavatta, and Marco Bellini \\ 
\end{center}
\end{widetext}
\setcounter{equation}{0}
\setcounter{page}{1}

\section{Mathematical properties of the PNQD}

By definition, the process-nonclassicality quasiprobability distribution (PNQD) is given by
\begin{align}
	P_{\Omega,\mathcal{E}}(\beta|\alpha) &=\frac{1}{\pi^2} \int d^2\xi
\Omega_{w}(\xi)  e^{\beta\xi^{*}-\xi\beta^{*}}\nonumber\\
	&\qquad \times\text{Tr}[e^{\hat{a}^{\dagger}\xi}e^{-\hat{a}\xi^{*}} \mathcal{E}(\ket{\alpha}\bra{\alpha})].
\end{align}
By using the Kraus decomposition of the quantum process, $\mathcal{E}(\rho) = \sum_{i=1}^{L}\hat{K}_{i}^{}
\, \rho \,\hat{K}_{i}^{\dagger}$, where $L\leq \text{dim}(\mathcal{H})^2$, $\sum_{i=1}^{L}\hat{K}_{i}^{}
\,\hat{K}_{i}^{\dagger}\leq\mathcal{I}$ and $\hat{K}_{i}$ are some Kraus operators on $\mathcal{H}$,
the above equation becomes
\begin{align}
P_{\Omega,\mathcal{E}}(\beta|\alpha)&=
\frac{1}{\pi^2} \sum_{i=1}^{L}\int d^2\xi
\Omega_{w}(\xi) e^{\beta\xi^{*}-\xi\beta^{*}}\nonumber\\
	&\qquad
\times\bra{\alpha}\hat{K}_{i}^{\dagger}e^{\hat{a}^{\dagger}\xi}e^{-\hat{a}\xi^{*}} \hat{K}_{i}^{}  \ket{\alpha}.
\end{align}
 Consequently, we can write the PNQD as an expectation value
\begin{equation}
	P_{\Omega,\mathcal{E}}(\beta|\alpha) = \bra \alpha\mathcal{E}_*(\hat O(\beta))\ket\alpha
\end{equation}
with respect to the input coherent state, where we have defined
\begin{equation}
	\mathcal{E}_*(\hat O_w(\beta)) = \sum_{i=1}^L \hat K_i^\dagger \hat O_w(\beta) \hat K_i
\end{equation}
and 
\begin{equation}
\hat O_w(\beta)\equiv \frac{1}{\pi^2} \int d^2\xi
\Omega_{w}(\xi)  e^{\hat a^{\dagger}\xi}e^{-\hat a\xi^{*}}  e^{\beta\xi^{*}-\xi\beta^{*}}.
\end{equation} 
 It has been shown in~\cite{UniNonclWitness} that $\hat O_w(\beta)$ is bounded, and ${\rm Tr}(\hat O_w(\beta)) = \pi^{-1}$. Therefore, also the operator $\mathcal{E}_*(\hat O_w(\beta))$ is bounded~\cite{Kraus}.
According to~\cite{Cahill,MW}, this implies that this function can be expressed as an everywhere convergent power series in terms of $\alpha$ and $\alpha^\ast$. Hence, for any unknown quantum process $\mathcal E$ and any complex number $\beta$, $P_{\Omega,\mathcal{E}}(\beta|\alpha)$ is uniquely determined by any small compact set of input coherent states $\ket\alpha$.

\section{Determination of nonclassicality quasiprobability distributions}

\subsection{Phase-sensitive nonclassicality quasiprobability distribution}

We reconstruct the nonclassicality quasiprobability distribution (NQD) $P_\Omega(\beta)$ from the quadrature distributions with the help of the relation
\begin{equation}
	P_\Omega(\beta) = \int_{-\infty}^\infty dx \int_0^\pi\frac{d\varphi}{\pi} p(x;\varphi) f_\Omega(x, \varphi; \beta, w).
\end{equation}
The pattern function $f_\Omega(x,\varphi; \beta, w)$ is given by
\begin{align}
	f_\Omega(x,\varphi; \beta, w) =& \int_{-\infty}^\infty db\,\frac{|b|}{\pi} e^{i b x}  e^{2i|\beta| b \sin(\arg(\beta)-\varphi-\tfrac{\pi}{2})} \nonumber\\ 
	&\times e^{b^2/2}\Omega_w(b)\nonumber ,  
\end{align}
where $\Omega_w(b)$ is the nonclassicality filter. In practice, the nonclassicality quasiprobability can be estimated from quadrature-phase value pairs $(x_i, \varphi_i)_{i=1}^N$ as
\begin{equation}
	P_\Omega(\beta) = \frac{1}{N} \sum_{i=1}^N f_\Omega(x_i,\varphi_i; \beta, w).
\end{equation}

 The data consists of $266000$ points taken at ten different phase values. To avoid certain computational artifacts, the evaluation is performed as described in the Supplementary Material of \cite{NonclQuasiProbSqueeze}.

As the quantum efficiency slightly differs for each state, we decided to remove its effect and show the results for the states with quantum efficiency $\eta = 1$. This can be achieved by some simple rescaling:
\begin{equation}
	P_\Omega(\beta; \eta = 1, w) = \eta P_{\Omega} (\sqrt{\eta}\beta;\eta, w/\sqrt{\eta}).\label{eq:remove:eta}
\end{equation}
On the left side of the equation, we have the nonclassicality quasiprobability of the ideal state. Therefore, it is sufficient to sample the nonclassicality quasiprobability of the lossy state with rescaled width and $\beta$. For details, see also \cite{NonclQuasiProbSPATS, NonclWitnessSimulation}.

\subsection{Phase-randomized nonclassicality quasiprobability distribution}

If we want to predict the outcome of a phase-insensitive quantum process for a phase-insensitive input quantum state (like a thermal state), it is sufficient to examine only the phase-randomized output states of the process. In general, however, even a classical process can lead to a phase-sensitive output state from a phase-insensitive input state, for example by coherent displacement. For the phase-randomized nonclassicality quasiprobability,
\begin{equation}
	\bar P_\Omega(a) = \frac{1}{2\pi}\int_0^{2\pi}d\phi P_\Omega(a e^{i\varphi}), 
\end{equation}
we can use the phase-randomized pattern function
\begin{eqnarray*}
	\bar f_\Omega(x; a, w) &=& \frac{1}{2\pi} \int_0^{2\pi}d\phi f_\Omega(x, \varphi; a e^{i\phi}, w)\\
	&=& \int_{-\infty}^\infty db\,\frac{|b|}{\pi} e^{i b x}  J_0(2a b) e^{b^2/2}\Omega_w(b).
\end{eqnarray*}
As it can be seen, the phase argument of the pattern function disappears. 

\section{Prediction of output state of the single-photon addition process for an input thermal state}

There is one critical point in the consideration of probabilistic quantum processes, such as the single-photon addition process, see \cite{TomoCreation}: The output of the quantum process formalism is given by
\begin{equation}
	\hat\rho_{\rm out} \propto \mathcal E_{\text{add}}(\hat \rho_{\rm in}) = \hat a^\dagger \hat\rho_{\rm in}\hat a.
\end{equation}
However, for probabilistic quantum processes the term on the right side is not a valid quantum state, since the density matrix is not correctly normalized. Therefore, we do not observe the right side directly, but its quantum state 
\begin{equation}
	\hat\rho_{\rm out} = \frac{\hat a^\dagger \hat\rho_{\rm in}\hat a}{{\rm Tr}(\hat a^\dagger \hat\rho_{\rm in}\hat a)} = \frac{\hat a^\dagger \hat\rho_{\rm in}\hat a}{1 + \langle\hat a^\dagger\hat a\rangle_{\rm in}},
\end{equation}
Therefore, if we insert the $P$ representation of the quantum state 
\begin{equation}
	\hat \rho_{\rm in} = \int d^2\alpha P_{\rm in}(\alpha) |\alpha\rangle\langle\alpha|,
\end{equation}
and the output state for input coherent states
\begin{equation}
	\hat\rho_{\rm out}(\alpha) = \frac{\hat a^\dagger\ket\alpha\bra\alpha\hat a}{1+|\alpha|^2},
\end{equation}
we find that 
\begin{eqnarray}
	\hat \rho_{\rm out} &=& \frac{1}{1 + {\rm Tr}(\hat\rho_{\rm in} \hat a^\dagger\hat a)} \int d^2\alpha P_{\rm in}(\alpha) \mathcal E_{\text{add}} (|\alpha\rangle\langle\alpha|) \nonumber\\
	&=& 
	 \frac{1}{1 + \langle\hat a^\dagger\hat a\rangle_{\rm in}} \int d^2\alpha P_{\rm in}(\alpha) (1+|\alpha|^2)\hat\rho_{\rm out}(\alpha),
\end{eqnarray}
where $\langle\hat a^\dagger\hat a\rangle_{\rm in} $ is the mean photon number of the input state for which we want to predict the output.

So far, we have sampled the nonclassicality quasiprobabilities of the output states $\hat\rho_{\rm out}(\alpha)$ for different coherent input states, $P_{\Omega,\mathcal{E}}(\beta|\alpha)$. From these, we can predict the NQD of the output state for an input state as 
\begin{equation}
	P_\Omega(\beta) = \frac{1}{1 + \langle\hat a^\dagger\hat a\rangle_{\rm in}}\int d^2\alpha P_{\Omega,\mathcal{E}}(\beta|\alpha) P_{\rm in}(\alpha)(1+|\alpha|^2),
\end{equation}	
where $P_{\rm in}(\alpha)$ is the $P$ function of the input state. If the latter is independent of the phase, the integral simplifies to 
\begin{equation}
	P_\Omega(\beta) = \frac{2\pi}{1 + \langle\hat a^\dagger\hat a\rangle_{\rm in}} \int_0^{\infty} d a\  a \bar P_{\Omega,\mathcal{E}}(\beta|a) P_{\rm in}(a)(1+a^2),
\end{equation}
where $ \bar P_{\Omega,\mathcal{E}}(\beta|a)$ is the phase-randomized PNQD described above. In practice, we evaluate this integral from the final number of measured input coherent state with the trapezoidal rule.
 A systematic error is estimated by the comparison of the result with the integration over a cubic spline interpolation function.

\end{document}